\documentclass[aps,prb,twocolumn,showpacs,groupedaddress,amsmath,amssymb,floatfix]{revtex4-1}
\usepackage[ansinew]{inputenc}
\usepackage{multirow}
\usepackage{mathrsfs}
\usepackage{revsymb}
\usepackage{slashed}
\usepackage{longtable}
\usepackage[T2A]{fontenc}
\usepackage{amsmath}
\usepackage{amssymb}
\usepackage{bm}
\usepackage{epsfig}
\usepackage{graphicx}
\usepackage{color}

\begin{document}

\title{Parity violation effects in the Josephson junction of a $p$-wave superconductor}

\author{N.~A.~Belov}
\affiliation{Max Planck Institute for Nuclear Physics, Saupfercheckweg 1, 69117 Heidelberg, Germany}
\author{Z.~Harman}
\affiliation{Max Planck Institute for Nuclear Physics, Saupfercheckweg 1, 69117 Heidelberg, Germany}

\begin{abstract}

The phenomenon of the parity violation due to weak interaction may be studied with superconducting systems.
Previous research considered the case of conventional superconductors. We here theoretically
investigate the parity violation effect in an unconventional $p$-wave ferromagnetic superconductor, and find that its
magnitude can be increased by three orders of magnitude, as compared to results of earlier studies.
For potential experimental observations, the superconductor UGe$_2$ is suggested, together with the description of a possible experimental scheme
allowing one to effectively measure and control the phenomenon. Furthermore, we put forward a setup for a further significant enhancement
of the signature of parity violation in the system considered.

\end{abstract}

\date{\today}

\pacs{74.20.Rp,74.90.+n,11.30.Er,12.15.-y}

% 74.20.Rp Pairing symmetries (other than s-wave)
% 74.90.+n Other topics in superconductivity
% 11.30.Er Charge conjugation, parity, time reversal, and other discrete symmetries
% 12.15.-y Electroweak interactions

\maketitle

\section{Introduction}

The electroweak theory, combining two fundamental interactions -- the electromagnetic and weak forces --
was introduced by Salam, Glashow and Weinberg in the 1960s~\cite{Sal64, Gla63, Wei67}. It explains the
nuclear beta-decay and weak effects in high-energy physics. One of the most prominent properties of the
electroweak theory is the spatial parity violation (PV). This unique phenomenon distinguishes the weak
interaction from the electromagnetic one, therefore, it helps to investigate weak properties
on an electromagnetic background. Firstly, PV was experimentally detected in the beta decay of $^{60}$Co
by Wu~\cite{Wu57} and collaborators. Later, many other novel experiments for the PV observation
have been proposed and performed. Low-energy PV experiments in atomic physics were carried out with Cs atoms
(see e.g. \cite{BB97,Woo97,Ben99}).
The PV effect has been theoretically predicted to have a measurable influence on the vibrational spectrum of
molecules in Ref.~\cite{Fag03}. Investigations of PV effects enable tests of
the standard model of elementary particle physics and impose constraints on physics beyond this model.
The search for new efficient ways to re-examine and investigate the PV phenomenon is an ongoing research activity
(see, for instance,~\cite{Sno11,Sha10,Sap03,Lab01}).

Another physical situation where PV effects can play a noticeable role is the interaction of electrons with the crystal lattice
of nuclei in the solid state~\cite{Khr74,Zhi82}. While the relative contribution of the PV effect is lower in comparison to other
investigation methods, PV experiments with solids are of interest because of the compact size of the experimental equipment.
Possible solid-state systems where one may study the PV contribution are superconductors (SC). Such systems would enable to study
the macroscopic manifestation of a quantum effect such as the electroweak interaction. The idea that PV effects
can appear in SCs has been suggested by Vainstein and Khriplovich~\cite{Khr74}. They have realized that the
electroweak contribution is insignificantly small in conventional $s$-wave SCs. However, it was predicted~\cite{Khr91},
that the effect can be increased by using SCs of other types, e.g. $p$-wave SCs.
Nowadays different unconventional SCs can be created and are well understood~\cite{IntroUSC}. Therefore,
we estimate this effect to be observed in $p$-wave ferromagnetic SCs, and put forward a novel method
for its observation and control. Our calculations yield a relative contribution of PV enhanced by several orders of
magnitude as compared to the $s$-wave case. The PV effect may still not be strong enough to be immediately
measurable, however, the results of this manuscript open the way for further enhancements of the signature of PV.

\begin{figure}[t]
   \center
   \includegraphics[width= 0.9 \columnwidth]{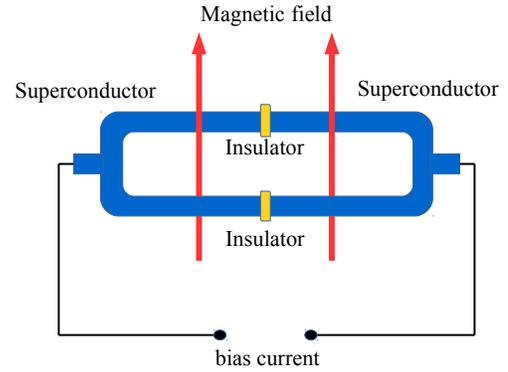}
   \caption{Circular Josephson junction with two insulations, placed into a magnetic field.}
   \label{fig:JJ}
\end{figure}

The SC system for studying PV considered in the present work is a circular Josephson junction in an external magnetic field.
This system is described, e.g., in Ref.~\cite{Fey63} and consists of a circular SC with two insulating junctions (see Fig.~\ref{fig:JJ}).
If the circular Josephson junction is placed in an external magnetic field, the maximal value of the SC current depends periodically
on the magnetic flux through the ring. This dependence is symmetric under the reflection of the direction of the magnetic flux.
However, the presence of the PV terms in the electron-nucleus interaction breaks this symmetry. We investigate this effect in
the case of unconventional $p$-wave SCs.

This work is organized as follows. In Section II we discuss the description of the PV effect in solid state (II.~A),
and present the possible superconducting system, namely, a circular Josephson junction (II.~B). In Section III,
we develop a method for the possible observation and control of the PV effect (III.~A), and evaluate the magnitude of
the PV effect for a certain SC, namely, uranium digermanide, UGe$_2$ (III.~B). In Subsection III.~C, we construct and
solve the equations for the coexistence of the ferromagnetic and superconducting phases. Section IV
discusses a scheme for a further improvement of the measurement technique, which can lead to a significant enhancement
of the PV effect. Finally, we provide a quantitative prediction of the PV effect in the system considered.

\section{Parity violation in superconductors}

\subsection{Weak odd-parity interaction in superconductors}

The weak odd-parity interaction in a crystal is given by the operator\cite{Khr74}
\begin{eqnarray}
W(\vec{r}) &=& \frac{G}{2\sqrt{2}m} \sum_i \bigg[ Zq\vec{\sigma}\left(\vec{p}\delta(\vec{r}-\vec{R_i})+
\delta(\vec{r}-\vec{R_i})\vec{p}\right)\nonumber \\
&+& \kappa_I \vec{I_i}\left(\vec{p} \delta(\vec{r}-\vec{R_i})+ \delta(\vec{r}-\vec{R_i})\vec{p}\right) \\
&+& i\kappa_I \vec{I_i} \vec{\sigma}
\left(\vec{p}\delta(\vec{r}-\vec{R_i})+\delta(\vec{r}-\vec{R_i})\vec{p}\right) \bigg]\,,\nonumber
\end{eqnarray}
where $G$ is the Fermi constant~\cite{CODATA2012}, $Z$ is the nuclear charge of crystal ions, $m$ stands for the electron
mass, $q$ is the weak charge, and $\vec{p}$ denotes the momentum of a Cooper pair. Furthermore, $I_i$ stands for the nuclear
spin, $\kappa_I$ is the pre-factor of the weak interaction of electrons with nuclear spins~\cite{Khr74}. The summation
goes over the crystal sites, determined by the position vectors $\vec{R}_i$. It is obvious from this equation that if
the pair spin is zero, $\sigma=0$ ($s$- or $d$-wave), only the second term is non-vanishing. This case has been investigated before in
Ref.~\cite{Khr91}. In the $p$-wave case~\cite{Buc81}, the first term is also nonzero, and its approximate magnitude is $Z$-times
larger than that of the other terms.

Now, we describe our case of interest ($\sigma=1$) in analogy with the $\sigma=0$ case~\cite{Khr91}. The first term of
the effective interaction is~\cite{Khr91}
\begin{equation}
\label{eq:Weff}
W_{\rm eff}^{(1)}=\frac{GZ^3qR}{\sqrt{2}}\frac{N}{2m}
\left[\vec{p}\vec{\sigma} + \vec{\sigma}\vec{p}\right]\,,
\end{equation}
where the weak charge is
\begin{equation}
q=\kappa_{1p}+\frac{A-Z}{Z}\kappa_{1n}\,,
\end{equation}
expressed with the factors $\kappa_{1p}=\frac{1}{2}(1-4\sin^2\theta_{\rm C})$ and  $\kappa_{1n}=-\frac{1}{2}$, where the Cabibbo weak mixing
angle~\cite{Cab63} is given by $\sin^2\theta_{\rm C}=0.22529$. In the above equation, $N$ and $A$ are the density and mass number of nuclei,
respectively, and
\begin{equation}
R=4\left(\frac{a_{\rm B}}{2Zr_0}\right)^{2-2\gamma} \bigg/ \Gamma^2(2\gamma+1)
\end{equation}
is the enhancement factor of relativistic effects at small distances~\cite{Khr91}, where $\gamma=1-\alpha^2Z^2/2$,
$r_0  \approx 1.2\cdot A^{1/3}$~fm is the nuclear radius, $a_{\rm B}$ denotes the Bohr radius, and $\Gamma$ denotes the gamma
function of real argument. $R$ is on the order of 10 for heavy elements.
The effective term (\ref{eq:Weff}) has to be added to the standard electromagnetic Lagrangian:
\begin{equation}
L=-m \sqrt{1-{v^2}} + e\vec{A}\vec{v}-e\phi-W_{\rm eff}^{(1)},
\end{equation}
where we use relativistic units. For the momentum of an electron one obtains
\begin{equation}
\vec{P}=\frac{\partial{L}}{\partial{\vec{v}}}= \frac{m\vec{v}}{\sqrt{1-v^2}}+e\vec{A}-
\frac{GZ^3qRN}{\sqrt{2}2m}2m^*\vec{\sigma}\,,
\end{equation}
where $m^*$ is the effective mass of the electron. The mass ration $\frac{m^*}{m}$ can exceed $10^2$.
The above weak modification of the electron momentum is equivalent to the substitution
\begin{equation}
e\vec{A}\rightarrow e\vec{A}-\frac{GZ^3qRN}{\sqrt{2}} \frac{m^*}{m}\vec{\sigma} \,,
\end{equation}
to be performed in all equations. We apply this substitution in the description of a superconducting ring.

Considering the case when the external magnetic field does not penetrate into the SC, one can obtain the following expression
for the magnetic flux in the superconducting ring:
\begin{equation}
2e\Phi\rightarrow 2e\Phi-f,
\label{eqn:PV_Phi2f}
\end{equation}
with $\Phi=\oint d\vec{r}\vec{A}$ and
\begin{equation}
f=\sqrt{2}GZ^3qRN\frac{m^*}{m}\oint d\vec{r}
\vec{\sigma}.
\label{eqn:PV_f}
\end{equation}
This result can be used in any applications and for any SC systems. In the following Section we apply this for the circular Josephson junction.
Let us discuss now the loop integral in Eq.~\ref{eqn:PV_f}. In the case of $p$-waves, the pairing spin is $\sigma=1$. However, the integral
\begin{equation}
\oint d\vec{r}\vec{\sigma}(\vec{r})
\end{equation}
is non-zero only if the mean spin vector (averaged over the whole SC circle) is non-zero. Therefore, it is advisable
to use an unconventional SC, which possesses a superconducting phase in coexistence with the ferromagnetic
phase. It allows one to control the effect by inducing magnetization in the SC. Furthermore, because of the $Z^3$ scaling of
the weak flux $f$, it is advantageous to employ heavy-element SCs. A possible material with these properties
is uranium digermanide, UGe$_2$~\cite{Sax00, Zeg12, Sho05}.

\subsection{Circular Josephson junction}

Our suggested experimental setup for the observation of the PV effect in SC is a circular Josephson junction (JJ).
A linear JJ is created by two SCs, separated by a thin insulator material.
It was predicted by Josephson \cite{Jos62} that the insulator does not prevent the appearance of a
superconducting current, however, the properties of the current depend on the thickness and material of this insulator.
Nowadays JJs have a wide spectrum of applications connected with atomic physics and quantum optics~\cite{You11}.
We consider the point-contact limit of the JJ, i.e. we assume that the insulator is infinitely thin.
However, all derivations presented here can be easily extended for other JJ models, since the PV effect
breaks the symmetry in any case.

The current in the JJ in the point contact approximation is given by the expression~\cite{Gol04}
\begin{equation}
J(\phi)=J_0\Delta \frac
{\tanh\left(\frac{\Delta}{2\pi}\sqrt{1-D\sin^2(\phi/2)}\right)}
{\sqrt{1-D\sin^2(\phi/2)}}
\sin\phi\,,
\end{equation}
where $D$ is the angle-averaged transmission probability, and $\Delta$ stands for the gap parameter. The phase is
defined by
\begin{equation}
\phi=\delta_0+2e\int A ds\,,
\end{equation}
where the integral is to be taken across the junction~\cite{Fey63} and $\delta_0$ is an unknown constant phase.
This expression is valid both in the clean and dirty limits of the SC.

\begin{figure}[t]
   \center
   \includegraphics[width=0.55 \columnwidth]{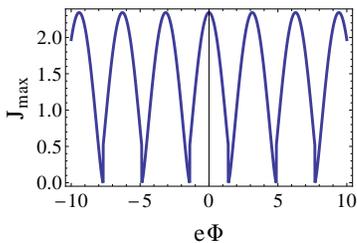}
   \caption{The well-known dependence of the maximal current
   $J_{\rm max}$ (in units of $J_0$) on the magnetic flux
   $e\Phi$ (in units of $\hbar$), with values $\Delta=100$ and $D=0.5$. $J_{\rm max}$ is calculated according
   to Eq.~(\ref{eqn:PV_Jtot}). Both energy and temperature are measured in units of Kelvin.
   See e.g. Ref.~\cite{Fey63} on the discussion of this function.}
   \label{fig:Jmax}
\end{figure}

If one now constructs a circular JJ by two identical JJ $a$ and $b$ connected in parallel (see Fig.~\ref{fig:JJ}),
only the following phase difference between these junctions is observable:
\begin{equation}
\delta_b-\delta_a=2e\oint Ads\,,
\end{equation}
where the circular integral is to be taken along the loop, and thus $\delta_b-\delta_a=2e\Phi$.
As noted above, we can only control the phase difference, thus one can write, following Ref.~\cite{Fey63}:
$\delta_a=\delta_0+e\Phi$ and $\delta_a=\delta_0-e\Phi$. The total current in the circular JJ as a
function of the magnetic flux is then given by the expression
\begin{eqnarray}
\label{eqn:PV_Jtot}
&&J_{\rm tot}(\Phi)=J_0\Delta \\
&&\times\left[
\frac
{\tanh\left(\frac{\Delta}{2\pi}\sqrt{1-D\sin^2((\delta_0+e\Phi)/2)}\right)}
{\sqrt{1-D\sin^2((\delta_0+e\Phi)/2)}}
\sin(\delta_0+e\Phi) \right. \nonumber \\
&&+\left.\frac
{\tanh\left(\frac{\Delta}{2\pi}\sqrt{1-D\sin^2((\delta_0-e\Phi)/2)}\right)}
{\sqrt{1-D\sin^2((\delta_0-e\Phi)/2)}}
\sin(\delta_0-e\Phi) \right]\,.\nonumber
\end{eqnarray}
This expression still depends on the arbitrary phase $\delta_0$. One may however determine the maximal value of the
current $J_{\rm max}$. The behavior of the maximal current can be calculated numerically for a certain gap parameter
$\Delta$ and a diffusion parameter $D$. The dependence of $J_{\rm max}$ on $e\Phi$ is shown in Fig.~\ref{fig:Jmax}
for the values $\Delta=100$ (in units of temperature) and $D=0.5$.

The dependence of $J_{\rm max}$ on $\Phi$ is invariant under the change of the sign of $\Phi$.
Due to the presence of the the weak odd-parity interaction, as shown in the previous Section,
one has to substitute $\Phi$ in all equations as
\begin{equation}
\label{eq:flux-subs}
2e\Phi\rightarrow 2e\Phi-f\,,
\end{equation}
where $f$ is the positive admixture of the weak parity-violating term. Thus the real dependence of the maximal
current on the magnetic flux is given by
\begin{equation}
J_{\rm max}^{\rm real}(\Phi)=J_{\rm max}(e\Phi-f/2)\,,
\end{equation}
and it is not symmetric with respect to the change of the sign of $\Phi$. The main purpose of this work
is to present the case in which this asymmetry can be measured. In the following Sections we discuss the
calculation and a possible measurement method for the value of $f$.

\section{A possible method for the measurement of the $f$-parameter}

\subsection{Method}

\begin{figure}[t]
   \center
   \includegraphics[width=0.55 \columnwidth]{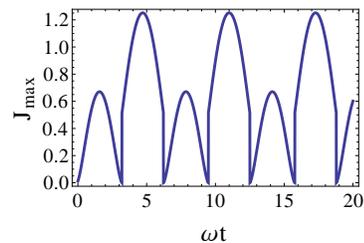}
   \caption{The periodic dependence
   $J_{\rm max}(\omega t)$ without the inclusion of the weak interaction.}
   \label{fig:Jper1}
\end{figure}

It can be challenging to directly observe the small asymmetry of the dependence of $J_{\rm max}^{\rm real}$ on $e\Phi$, therefore,
we suggest to employ a time-periodic magnetic field instead of a static one. As mentioned before, one can compute the phase-independent
maximal Josephson current $J_{\rm max}(e\Phi)$. Let $e\Phi_0$ be the first positive root of this expression. 
If we now introduce a periodic component to the magnetic field,
\begin{equation}
e\Phi(t)=e\Phi_0+\frac{a}{2}\sin{\omega t}\,,
\end{equation}
$J_{\rm max}(\omega t)$ also depends periodically on time with the period $T=2\pi/\omega$, where $\omega$ is the angular frequency
of the oscillating field. Roots of this function are reached every half of the period, i.e. with a $\pi/\omega$ periodicity. 
The typical shape of this function, calculated for the case of $J_{\rm max}(e\Phi)$ shown on Fig. \ref{fig:Jmax},
is presented on Fig. \ref{fig:Jper1}.

Let us now incorporate the weak interaction into this system. One can see that the weak interaction can be controlled
("switched on/off") by introducing the magnetization in our SC ferromagnetic circular JJ, since the PV contribution is
proportional to the average spin of the Cooper pairs [see the integral of the spin over a circle in Eq.~(\ref{eqn:PV_f})].
The periodic field coefficient $a$ is chosen to be greater than the weak factor $f$, however,
it is comparable to it: $a=x f,\, x\gtrapprox 1.$

\begin{figure}[t]
   \center
   \includegraphics[width=0.55 \columnwidth]{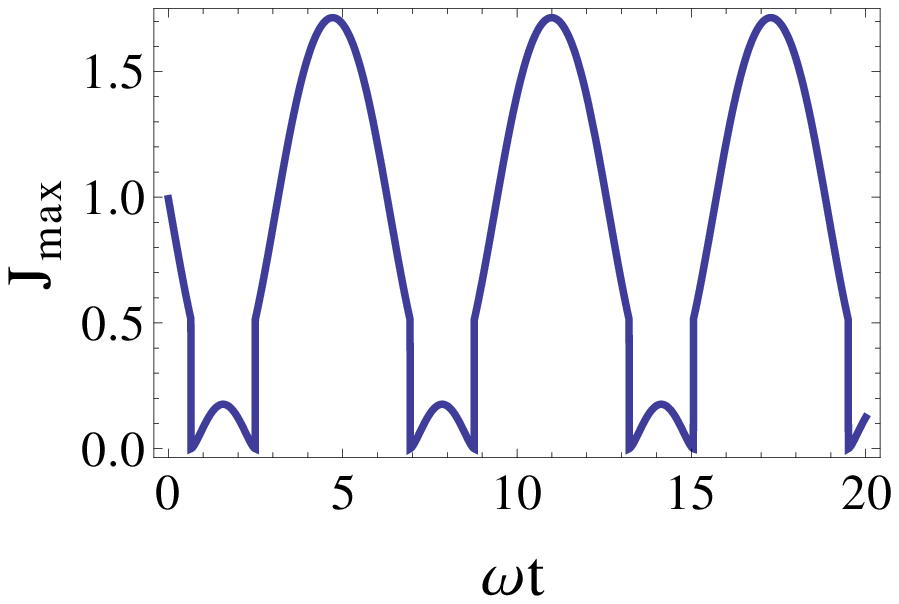}
   \includegraphics[width=0.55 \columnwidth]{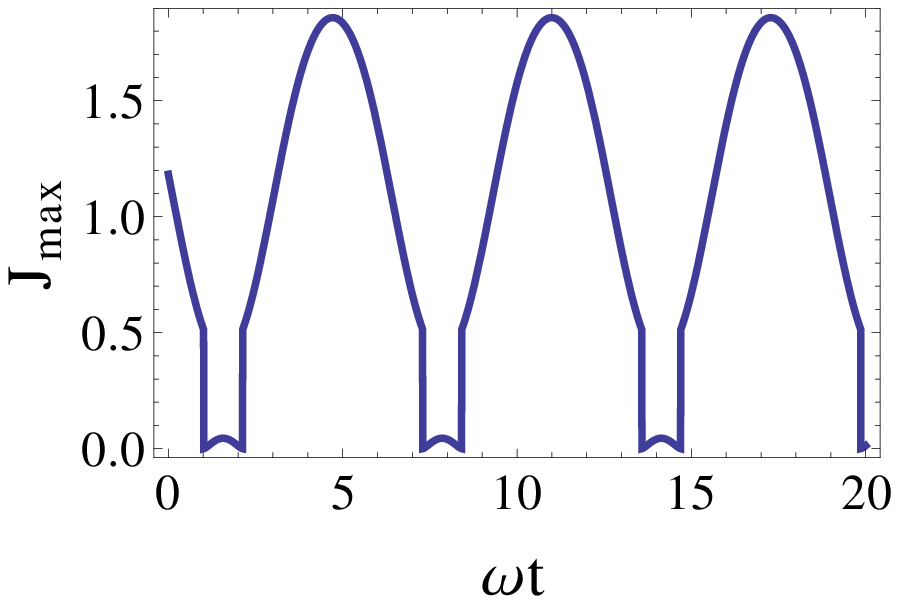}
   \includegraphics[width=0.55 \columnwidth]{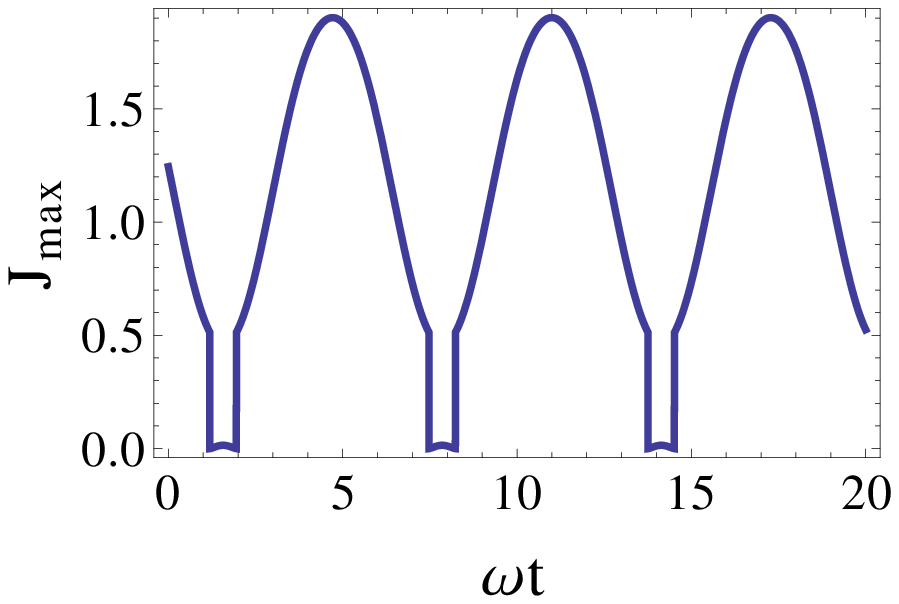}
   \caption{The periodic dependence
   $J_{\rm max}(\omega t)$ with the weak interaction included, for the parameters
   $a=xf$, $x=$1.5, 1.1, 1.01, respectively.}
   \label{fig:Jper2}
\end{figure}

Now the roots of $J_{\rm max}(\omega t)$ are not exactly $\pi/\omega$-periodic any more. This behavior is shown on Fig.~\ref{fig:Jper2}.
Furthermore, in the limit $x\rightarrow 1$, the roots become almost $2\pi/\omega$ - periodic.
This non-periodic behavior of roots can be observed experimentally, since it can be dynamically controlled
by the periodic magnetic field.

\begin{figure}[t]
   \center
   \includegraphics[width=0.55 \columnwidth]{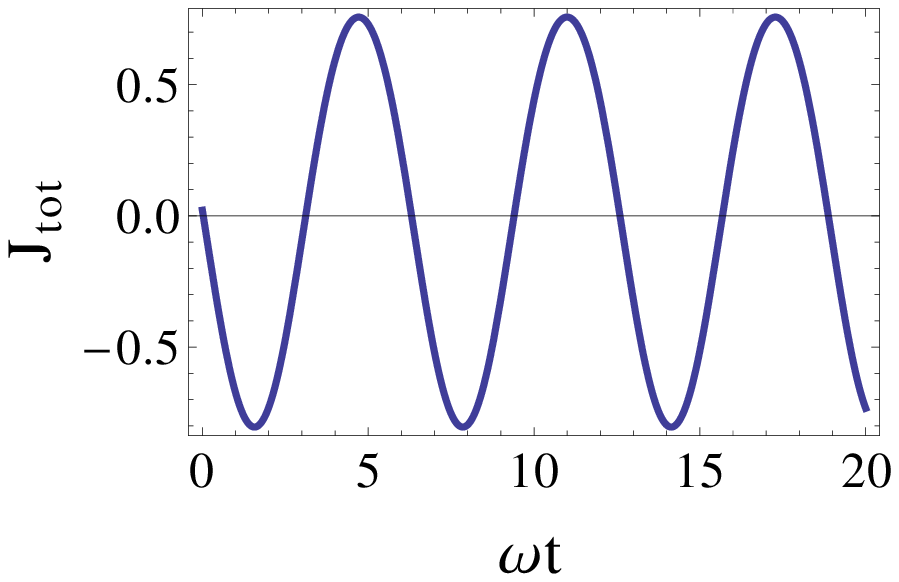}
   \includegraphics[width=0.55 \columnwidth]{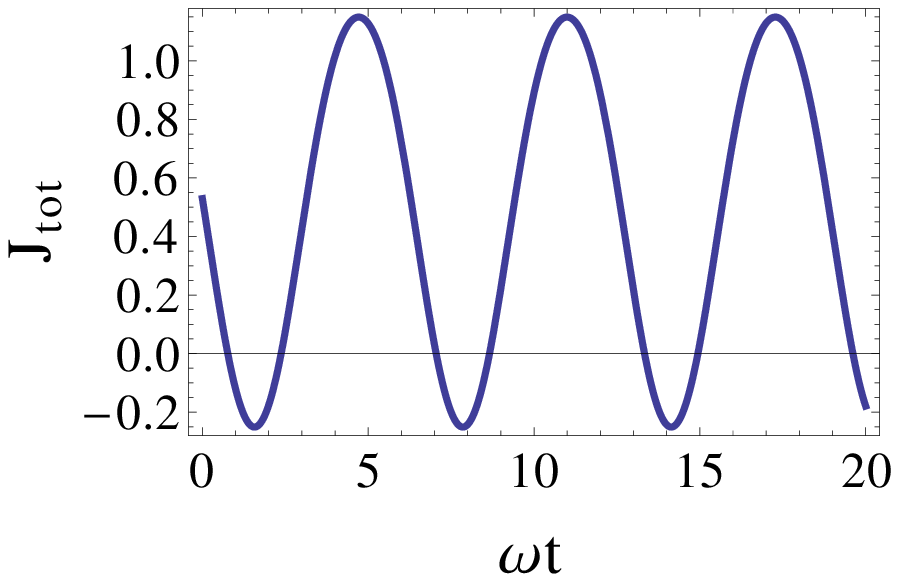}
   \includegraphics[width=0.55 \columnwidth]{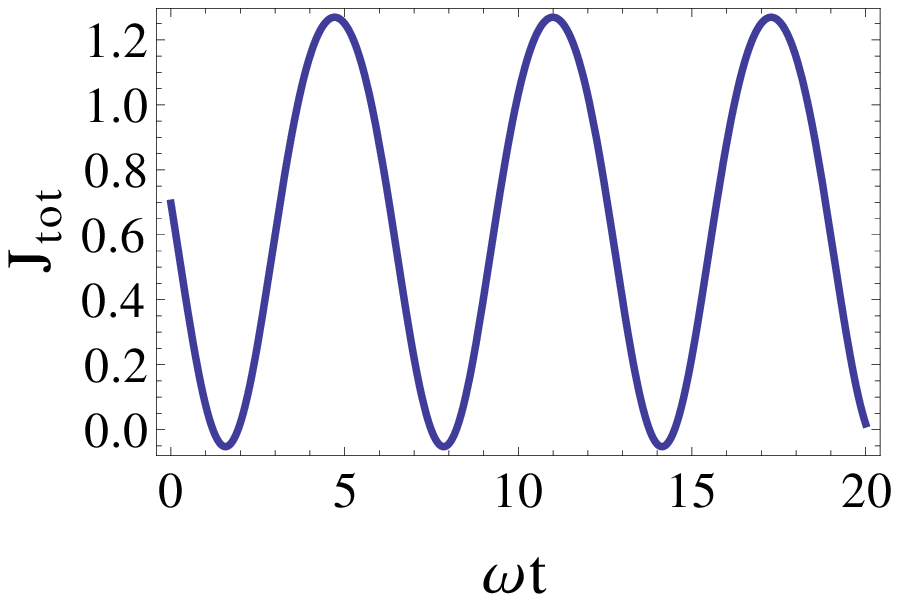}
   \caption{The periodic dependence
   $J_{\rm tot}(\delta_0,\omega t)$. The first plot shows the case when  the
   weak interaction is not included, and the next two plots are for the case when
   the weak interaction is included, with the parameters $a=xf$, $x=1.5,\,1.1$,
   respectively.}
   \label{fig:Jper3}
\end{figure}

As an alternative, one may also do the measurement at some certain phase $\delta_0$ rather than at
a maximal Josephson current. Here we can introduce again the oscillations of the magnetic field
around the first root $e\Phi_0$ of the total current $J_{\rm tot}(\delta_0,e\Phi)$.
In this case, the current changes its sign during the total period $T=2\pi/\omega$.
If we take $a=f$ after switching on the weak interaction, with magnetization being present in the SC,
the total current function will be always of the same sign, as it is shown on Fig.~\ref{fig:Jper3}.

\subsection{Estimation of the effect}

In the present Section we evaluate the value of the admixture $f$ to the magnetic flux through the JJ ring
[see Eqs.~(\ref{eqn:PV_Phi2f}),~(\ref{eqn:PV_f})] to provide an estimate of the magnitude of the PV effect in SCs.

In the case of a ferromagnetic SC we can assume that pairs are polarized along the loop, therefore,
their polarization can have two different opposite directions. This assumption yields for the loop integral
\begin{equation}
\oint d\vec{r}\vec{\sigma}(\vec{r}) = \eta\oint dr\,,
\end{equation}
with the mean spin value $\eta$, which has to be
determined by a self-consistent solution of the equations for superconductivity
and ferromagnetism in this material. This is performed in the following Section.

Now we give an estimation of the PV effect. The PV admixture $f$ is expressed with the mean spin $\eta$ as
\begin{equation}
f=\sqrt{2}GZ^3RqN\frac{m^*}{m}
\eta\oint dr\,.
\end{equation}
Assuming a round JJ, the loop integral simply yields $\oint dr=2\pi L$, the mean spin value
$\eta$ is to be calculated in the next section, and the remaining factors are known:
$\frac{G}{\pi\alpha^3}=10^{-13}$, $Z=92$. The relativistic enhancement parameter is $R\approx 11$,
the value of the effective mass  at the ambient pressure is in the interval~\cite{Suz94}
$\frac{m^*}{m}=2.3 \dots 25$, thus we may assume $\frac{m^*}{m}=25.$
The density of nuclei is~\cite{Bou97,SPD} $N=0.25\cdot 10^8~{\rm cm}^{-1}$.
Then, in dimensionless units ($\hbar=1$), the final value of $f$ is
\begin{equation}
\label{eq:approxf}
f=2\pi \cdot 6.9\cdot 10^{-4} L\eta,
\end{equation}
where the length $L$ is measured in units of cm. This result is 3 orders of magnitude larger than the value of the
admixture factor in the case of an $s$-wave heavy SC~\cite{Khr91}. To observe this effect one may use
the method with the oscillating magnetic field, as described in the previous Section.

Since the flux, in units of $\hbar$, is given by
\begin{equation}
e\Phi=\frac{e\pi L^2}{\hbar}B\,,
\end{equation}
the time-dependent part of the magnetic field is determined by
\begin{equation}
\label{eqn:flux}
e\Phi_0+\frac{a}{2}\sin{t}=\frac{e\pi L^2}{\hbar}(B_0+B_t\sin{t})\,.
\end{equation}
In case of $a \gtrapprox f$,  the expression for the amplitude $B_t$, in units of Tesla is [cf. Eq.~(\ref{eq:flux-subs})]
\begin{equation}
0.152\cdot 10^{20} L^2 B_t  \gtrapprox \frac{f}{2},
\end{equation}
where $L$ is given in cm. Inserting our estimate for $f$ [see Eq.~(\ref{eq:approxf})], it follows:
\begin{equation}
\label{eqn:PV_Bt1}
B_t=4.5\cdot 10^{-23} \frac{\eta}{L}\,[T].
\end{equation}
As an example, for a typical size of $L=0.1~{\rm \mu m}$, the result is $B_t=4.5\cdot 10^{-18}\eta$~[T].
We discuss this result in the concluding Section after showing in the following that $\eta$ can indeed
reach its maximal value, $\eta_{\rm max}=1$.

\subsection{Calculation of the mean spin value}

For the calculation of the mean spin value~$\eta$ of Cooper pairs we use a model for the coexistence
of superconductivity and ferromagnetism of Ref.~\cite{Jia09}, described there for the case of an isotropic material.
We extended this formalism for anisotropic materials. The analytical derivations are similar to those of Ref.~\cite{Jia09},
however, for completeness, we present them here, together with the description of the numerical algorithm used.

In this model, the Hamiltonian is given as
\begin{eqnarray}
H&=&\sum_{k \sigma}(\epsilon-\mu-\sigma M)c_{k \sigma}^+c_{k \sigma} \\
&&-\frac{1}{2\nu}\sum_{k k',\sigma \sigma'}V(kk')
c_{k \sigma}^+c_{-k \sigma'}^+c_{-k' \sigma'}c_{k' \sigma},\nonumber
\end{eqnarray}
where $\sigma=\pm1$ denotes single-particle spin states, $k$ is the single-particle momentum,
$\epsilon$ denotes the non-magnetic part of the quasi-particle energy, $c_{k\sigma}^+$ and $c_{k\sigma}$
are quasi-particle creation and annihilation operators, respectively. Furthermore, $\mu$ is the chemical
potential, $\nu$ is the sample volume, $V$ stands for the pairing potential, and the magnetization is
$M=U(n_+-n_-)/2$, defined in terms of the Stoner parameter $U$ and number of pairs with the spin in the
direction of the magnetization ($n_+$) and in the opposite direction ($n_-$). The Stoner parameter depends
on the pressure, but it is independent of the temperature. In the ferromagnetic phase, only the pairs with
spins parallel to the field can exist. We introduce two gap parameters $\Delta_\pm$ for spins in the direction
of magnetization $(+)$ and in the opposite direction $(-)$.

In Ref.~\cite{Jia09}, the Matsubara Green's functions \cite{AGD65} for this Hamiltonian are
constructed, and, after summation over Matsubara frequencies, the equations are obtained for
the magnetization, number of particles and gap parameters. By replacing all summations by
continuum integrals in dimensionless energy units, rescaled by the factor $\frac{\hbar^2}{2m^*}$,
one receives the equations
\begin{eqnarray}
\label{eqn:PV_a}
M&=&\frac{U}{64\pi^3}\int_0^\infty d\epsilon_0\int_0^\pi d\theta
\int_0^{2\pi}d\phi \sin(\theta)\sqrt{\epsilon_0} \\
&\times&\left(\frac{\epsilon_- \tanh(E_-/2T)}{E_-}-
\frac{\epsilon_+ \tanh(E_+/2T)}{E_+}\right),\nonumber \\
\label{eqn:PV_b1}
1&=&\frac{V}{64\pi^3}\int_{\epsilon_{F+}-\omega_+}^{\epsilon_{F+}+\omega_+}d\epsilon_0
\int_0^\pi d\theta\int_0^{2\pi}d\phi \\
&&\left(\frac{\sqrt{\epsilon_0}\sin^3\theta}{E_+}\tanh(E_+/2T)\right)\,,\nonumber \\
\label{eqn:PV_b2}
1&=&\frac{V}{64\pi^3}\int_{\epsilon_{F-}-\omega_-}^{\epsilon_{F-}+\omega_-}d\epsilon_0
\int_0^\pi d\theta\int_0^{2\pi}d\phi\\
&&\left(\frac{\sqrt{\epsilon_0}\sin^3\theta}{E_-}\tanh(E_-/2T)\right)\,,\nonumber\\
\label{eqn:PV_c}
1&=&\frac{1}{32\pi^3}\int_0^\infty d\epsilon_0\int_0^\pi d\theta \int_0^{2\pi}d\phi
\sin(\theta)\sqrt{\epsilon_0}\\
&\times&\left(2-\frac{\epsilon_- \tanh(E_-/2T)}{E_-}-
\frac{\epsilon_+ \tanh(E_+/2T)}{E_+}\right)\,,\nonumber
\end{eqnarray}
where the following quantities have been introduced: $\epsilon_{F\pm}=\mu\pm M$, $\epsilon_\pm=\epsilon-\epsilon_{F\pm}$,
$E_\pm=\sqrt{\epsilon_\pm^2+\sin^2\theta\Delta_\pm^2}$, and $\omega_\pm=0.01\epsilon_{F\pm}$.
The integration over the variables $\epsilon_0$, $\theta$ and $\phi$ corresponds
to an integration over the three-dimensional momentum of the pair.
Equation~(\ref{eqn:PV_a}) is the expression for the magnetization in the ferromagnetic SC.
Eq.~(\ref{eqn:PV_b1}), together with Eq.~(\ref{eqn:PV_b2}), presents the gap equation for pairs polarized
in or opposite to the direction of the magnetization. Finally, Eq.~(\ref{eqn:PV_c}) expresses the
conservation of the number of pairs.

In an isotropic case, considered before in Ref.~\cite{Jia09}, the relation
$\epsilon=\epsilon_0$ holds. However, for anisotropic materials, during the change of
the summation over $\vec{k}$ to three dimensional integration, the angular integrals
in spherical coordinates remain the same, however, the radial variables are changed:
\begin{eqnarray}
k^2=k_0^2\bigg(
\left(\frac{c}{a}\sin\theta \cos\phi\right)^2+\left(\frac{c}{b}\sin\theta \sin\phi\right)^2\\
+\left(\frac{c}{c}\cos\theta \right)^2\bigg)\,,\nonumber
\end{eqnarray}
with $a$, $b$, $c$ being the crystal cell parameters. Thus, we have 3 integrals over $k_0$, $\theta$ and $\phi$,
which change to integrals over $\epsilon_0$, $\theta$ and $\phi$, and the energy in all equations depends on the angles:
\begin{eqnarray}
\epsilon=\epsilon_0\bigg(\left(\frac{c}{a}\sin\theta \cos\phi\right)^2+\left(\frac{c}{b}\sin\theta \sin\phi\right)^2\\
+ \left(\frac{c}{c}\cos\theta \right)^2\bigg)\,.\nonumber
\end{eqnarray}
Finally, we arrive to 4 equations, Eqs.~(\ref{eqn:PV_a}-\ref{eqn:PV_c}), for 4 the variables $M,\Delta_\pm,\mu$, with
$U$, $V$ and $T$ as parameters. These equations have to be solved self-consistently. The sought-after mean spin
value is given by $\eta=2M/U$. The Stoner parameter $U$ is determined by the Curie temperature $T_{\rm C}$ at a certain
pressure \cite{Kit53}. We obtain it by a self-consistent solution of Eqs.~(\ref{eqn:PV_a}) and~(\ref{eqn:PV_c})
with $\delta\equiv0$ and assuming the condition that the magnetization appears at temperatures $T<T_{\rm C}$ only.
The method of this solution is similar to the method for the calculation of $\eta$ presented below. The pairing
parameter $V$ is determined by the condition that at temperatures below the critical SC temperature,
$T<T_{\rm sc}$, the following should hold: $\Delta\ne0$, and at $T>T_{\rm sc}$ there is no superconductivity ($\Delta=0$).

\begin{figure}[t]
\center
\includegraphics[width=0.55 \columnwidth]{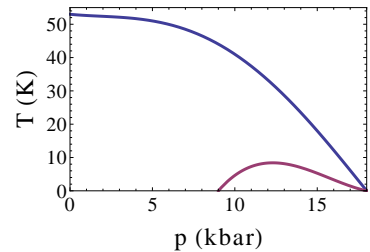}
\caption{The $p$--$T$ phase diagram for UGe$_2$~\cite{Hux01}.
The lower curve $10\times T_{\rm sc}(p)$ corresponds to the critical temperature of superconductivity,
the upper one to the Curie critical temperature of ferromagnetism, $T_{\rm C}(p)$.}
\label{fig:PV_Hux01}
\end{figure}

We use the following algorithm for the self-consistent solution of the full set of equations
[Eqs.~(\ref{eqn:PV_a})-(\ref{eqn:PV_c})] to evaluate $\eta$ for certain values of $U$ and $V$
(parameters of the SC): (i) With the help of Eq.~(\ref{eqn:PV_b1}) [or Eq.~(\ref{eqn:PV_b2})]
one can construct the function $\Delta(k)$ in such a way that $\Delta_\pm=\Delta(\mu\pm M)$. It is possible
to do so since both equations depend on the combinations $\mu\pm M$ only. (ii) By Eqs.~(\ref{eqn:PV_a})
and (\ref{eqn:PV_c}) we can construct the equations
\begin{equation}
M=\mp\frac{U}{2}W(\mu\pm M)\,,
\end{equation}
where
\begin{eqnarray}
W(k)&=&1-\frac{1}{16\pi^3}\int_0^\infty d\epsilon_0\int_0^\pi d\theta
\int_0^{2\pi}d\phi\sqrt{\epsilon_0}\sin\theta\nonumber \\
&&\times \left(1-\frac{\epsilon_\pm \tanh(E_\pm/2T)}{E_\pm}\right)\,.
\end{eqnarray}
Let us take $x=\mu-M$, yielding two simple equations, namely:
\begin{equation}
M=\frac{U}{2}W(x)\,, \quad
M=-\frac{U}{2}W(x+2M)\,,
\end{equation}
which deliver the final equation for $x$,
\begin{equation}
W(x)=-W(x+U W(x))\,.
\end{equation}
From $x$ we can obtain the values of all parameters as follows:
\begin{eqnarray}
M&=&\frac{U}{2}W(x)\,,  \quad \mu=x+M\,, \\
\Delta_-&=&\Delta(x)\,, \quad \Delta_+=\Delta(x+2M)\,. \nonumber
\end{eqnarray}
Let us note that these equations have a solution with $M\ne0$ in the case when $U>U_{\rm c}$ only.
$U_{\rm c}$ is the critical value of the Stoner parameter and it depends on $T$.

\begin{figure}[t]
\center
\includegraphics[width=0.8 \columnwidth]{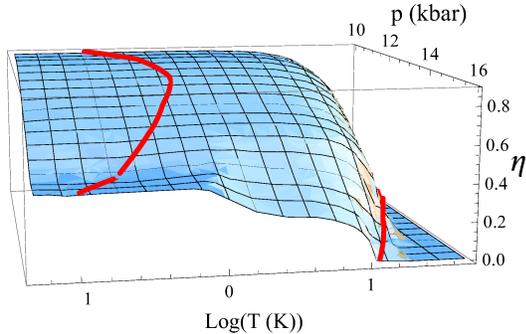}
\caption{The dependence  of the mean spin $\eta$ on pressure and temperature in the
regions $T<T_{\rm sc}$ (below the first red curve) and $T_{\rm sc}<T<T_{\rm C}$ (between the
red curves). The data on $T_{\rm sc}(p)$ and $T_{\rm C}(p)$ are taken from Ref.~\cite{Hux01}.}
\label{fig:PV_eta}
\end{figure}

Using this numerical algorithm we provide calculations for the cell parameters of UGe$_2$, namely,
$a=14.928$~pm, $b=4.116$~pm, $c=4.036$~pm \cite{Bou97}. We perform calculations for different pressures
and temperatures both in the region of the coexistence of ferromagnetic and superconductive
phases~\cite{Hux01} as well as in the pure ferromagnetic region (see Fig.~\ref{fig:PV_Hux01}).
It appears (see Fig.~\ref{fig:PV_eta}) that at all pressures between $9$ and approximately $12$~kbar,
the value of $\eta$ is almost unity for all temperatures $T<T_{\rm sc}$, however, above $~12$~kbar,
$\eta$ decrees with the increase of the pressure. These numerical results show that at some pressures
in the region of interest where $T_{\rm C}$ is much larger than $T_{\rm sc}$, $\eta$ is equal to unity.

\section{Possible experimental setup to increase the parity violation effect}

\begin{figure}[t!]
   \center
   \includegraphics[width=0.8 \columnwidth]{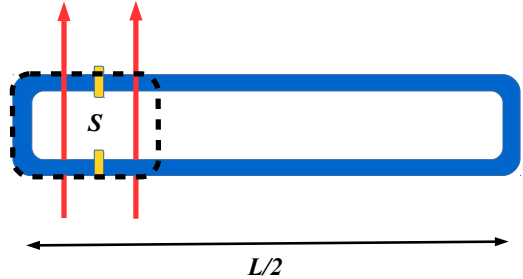}
   \caption{The circular Josephson junction with an external magnetic field present in
   the effective area $S$ in the vicinity of the SIS junctions only.}
   \label{fig:JJ_S}
\end{figure}

We assumed in the previous derivations that the induced magnetic field is constant along the loop.
However, the required periodic component of the magnetic field can be increased if the magnetic field
is only present in the region around the Josephson junctions (see Fig.~\ref{fig:JJ_S}). To make this
statement clear, we rewrite Eq.~(\ref{eqn:flux}) as
\begin{eqnarray}
\label{eqn:flux_all}
e\Phi_0+\frac{a}{2}\sin{t}&=&\int\int dS\frac{e}{\hbar}(B_0 + B_t \sin{t})\nonumber\\
&=&\frac{eS}{\hbar}(B_0+B_t\sin{t})\,,
\end{eqnarray}
where $S$ is the effective area (part of the loop area, see Fig.~\ref{fig:JJ_S}), where the field is
given by $(B_0+B_t\sin{t})$. Therefore, in the limit $a\rightarrow f+$, the expression for the periodic
component of the magnetic field is
\begin{equation}
\label{eqn:PV_Bt1_S}
B_t=1.4\cdot 10^{-22} \frac{\eta L}{S}~[{\rm T}]\,.
\end{equation}
This expression explains why the localization of the magnetic field by the decrease of the total flux
increases the required magnetic field. By choosing a large ratio $L/S$ one may reach conditions satisfying
the restrictions of existing experimental techniques. A large value of the ratio may be archived, e.g.,
by implementing a superconducting solenoid at the field-free part of the loop (i.e. at the right
side of  Fig.~\ref{fig:JJ_S}). Furthermore, the ratio $B_t/B_0\approx L$, therefore, the large size
of the loop provides also an improvement of the effect.

\section{Discussion and conclusions}

We have shown in Subsection III.C that the maximal mean spin value can be equal to unity, $\eta=1$,
for Cooper pairs in the unconventional ferromagnetic SC at some certain conditions,
namely, in the region where $T_{\rm sc} \ll T_{\rm C}$. Thus we can finally obtain the amplitude
of the periodic magnetic field required for the estimation presented in Subsection III.B
[cf. Eq.~(\ref{eqn:PV_Bt1})]:
\begin{equation}
B_t=4.5\cdot 10^{-23} \frac{1}{L}\,,
\end{equation}
where $L$ is given in units of cm and $B_t$ in units of Tesla. As an example, for the size of the circular
Josephson junction $L=0.1~{\rm\mu m}$, one obtains the value
\begin{equation}
B_t=4.5\cdot 10^{-18}\,[{\rm T}]\,.
\end{equation}
Thus, the PV effect is 3 orders of magnitude stronger than in the case of the earlier theoretical
works~\cite{Zhi82,Khr74}. These magnetic fields are close to the range of Superconducting
QUantum Interference Devices (SQUIDs, see e.g. Ref.~\cite{Sac14}).
The observation of PV might be disturbed by the appearance of spontaneous currents caused by broken time-reversal symmetry (see, e.g., Ref.~\cite{Luk98}).
Future research may explore effective ways for a further enhancement and control of the PV effect.

Furthermore, the effect can be significantly improved by employing the experimental scheme described
in Section IV. For instance, without the implementation of this model, the magnetic field is
\begin{eqnarray}
B_t&=&1.4\cdot 10^{-22}\frac{\eta L}{S}~[{\rm T}] \\
   &=&4.5\cdot 10^{-18}~[{\rm T}]|_{\eta=1,\,L=0.1\mu m,\,S=\pi L^2}\,; \nonumber
\end{eqnarray}
then, by increasing the length of the loop to $L=1$~mm at unchanged $S$, $B_t$ is augmented by 4 orders of magnitude:
\begin{equation}
B_t=4.5\cdot 10^{-14}[{\rm T}]|_{\eta=1,\,L   =1mm,\,S=\pi L_o^2,\,L_o=0.1\mu m}\,.
\end{equation}
Therefore, the PV effect is now 7 orders of magnitude larger than in the case of the earlier
proposals~\cite{Zhi82,Khr74}. As a result, we anticipate that PV effects in SC can be observed in future.
Such measurements will open the way to investigate the PV phenomenon by relatively compact experimental setups,
and offer one to study electroweak effects in a macroscopic system.

\begin{acknowledgements}

We acknowledge insightful conversations with Andreas Fleischmann, Christian Enss and Loredana Gastaldo.

\end{acknowledgements}

\end{document}